\documentstyle[12pt]{article}
\title{Stochastic stability in three-player games} 
\author{Dominik Kami\'{n}ski \\
Jacek Mi\c{e}kisz \\ Marcin Zaborowski
\\ Institute of Applied Mathematics and Mechanics 
\\ Warsaw University  \\ ul. Banacha 2  \\ 02-097
Warsaw, Poland} 
\begin{document} 
\baselineskip=26pt
\maketitle 

\noindent {\bf Abstract.} 
Animal behavior and evolution can often be described by game-theoretic models.
Although in many situations, the number of players is very large, 
their strategic interactions are usually decomposed into a sum of two-player games.
Only recently evolutionarily stable strategies were defined for multi-player games 
and their properties analyzed (Broom {\em et al.,} 1997). Here we study the long-run behavior 
of stochastic dynamics of populations of randomly matched individuals playing 
symmetric three-player games. We analyze stochastic stability of equilibria 
in games with multiple evolutionarily stable strategies. We also show that 
in some games, a population may not evolve in the long run to an evolutionarily stable equilibrium.
\eject

\section{Introduction}
Long-run behavior of interacting individuals can often be described within 
game-theoretic models. The basic notion here is that of a Nash equilibrium. 
This is a state of a population - an assignment of strategies to players - such that 
no player, for fixed strategies of his opponents, has an incentive to deviate from
his current strategy; the change cannot increase his payoff. Maynard Smith and Price (1973)
have refined this concept of equilibrium to include the stability of Nash equilibria against
mutants (see also Maynard Smith, 1982). 
They introduced the fundamental notion of an evolutionarily stable strategy. 
If everybody plays such a strategy, then the small number of mutants playing 
a different strategy is eliminated from the population. 
 
Although in many models, the number of players is very large, 
their strategic interactions are usually decomposed into a sum of two-player games.
However, truly multi-player games naturally appear in many situations.
For example, in the stag-hunt game, there are $n$ players who choose simultaneously 
one of two actions: either join the stag hunt (strategy $S$) or go after a hare (strategy $H$).
The strategy $H$ yields a safe but low fixed payoff. The strategy $S$  gives
the higher payoff if at least $k$ players join the stag hunt and split the reward.
The game has two Nash equilibria: the efficient one, where all players choose $S$ and 
the one in which players are averse to risks and play $H$. This is a typical situation
with a selection problem. The long-run outcome of such games is the subject of this paper.
In particular, we will show that in the stag-hunt game, strategy $S$ is played 
with a high frequency in the adaptive dynamics in the presence of mutations. 
 
Recently there have appeared some systematic studies of truly multi-player games. 
Broom {\em et al.} (1997) defined evolutionarily stable strategies for multi-player games 
and analyzed their properties. Kim (1996) investigated an asymptotic and stochastic stability 
of Nash equilibria in multi-player games.

Bukowski and Mi\c{e}kisz (2004) provided a classification of symmetric three-player games 
with two strategies. For certain payoff parameters, such games have multiple evolutionarily 
stable strategies. In one class of three-player games, there are two pure evolutionarily 
stable strategies like in the stag-hunt game. In another class, we have one pure and one mixed 
evolutionarily stable strategy. In both cases we are faced with a standard problem of equilibrium selection.
We will approach this problem from a dynamical point of view.

The dynamical interpretation of the evolutionarily stable strategy was provided 
by several authors (Taylor and Jonker, 1978; Hofbauer {\em et al.}, 1979; Zeeman, 1981). 
They proposed a system of differential or difference equations, the so-called 
replicator equations, which describe the time-evolution of frequencies of strategies. 
It is known that any evolutionarily stable strategy is an asymptotically 
stable stationary point of such dynamics (Weibull, 1995; Hofbauer and Sigmund, 1998 and 2003).   

Here we discuss a discrete-time stochastic adaptation dynamics 
of a population of a fixed number of individuals playing three-player 
games with two strategies. The selection part of the dynamics ensures that if the mean payoff 
of a given strategy at the time $t$ is bigger than the mean payoff of the other one, 
then the number of individuals playing the given strategy should increase in $t+1$. 
We assume that individuals are randomly matched with their opponents.
Such model was introduced and analyzed for the case of two-player games by Robson and Vega-Redondo (1996a, 1996b).
Alternatively, if the rate of playing games is much bigger than the rate of adaptation, then one may assume 
(as in the standard replicator dymanics) that individuals receive average payoffs 
with respect to all possible opponents - they play against the average strategy (Kandori {\em et al.}, 1993). 
In both models, players may mutate with a small probability hence the population may move against 
a selection pressure. To describe the long-run behavior of stochastic dynamics, 
Foster and Young (1990) introduced a concept of stochastic stability. 
A configuration of a system (a number of players playing 
the first strategy in our models) is stochastically stable if it has a positive probability 
in the stationary state of the dynamics in the limit of no mutations. 
It means that in the long run we observe it with a positive frequency. 
In the Kandori-Mailath-Rob model, the risk-dominant strategy (the one which has a higher expected
payoff against a player playing both strategies with equal probabilities) is stochastically stable - 
if the mutation level is sufficiently small, we observe it in the long run 
with the frequency close to one. In the model of Robson and Vega-Redondo, 
the payoff-dominant strategy (also called efficient) is stochastically stable. 
It is one of very few models in which a payoff-dominant strategy is stochastically stable 
in the presence of a risk-dominant one. 

Here we analyze the model of Robson and Vega-Redondo in the case of symmetric three-player games. 
We characterize stochastic stability of equilibria in all generic classes of such games. 
In particular, in games with multiple evolutionarily stable strategies 
we show that only one is stochastically stable and therefore we resolve 
the problem of equilibrium selection. We show that stochastic stability 
may depend on the number of players. We also show that in some games, 
a population may not evolve in the long run to an evolutionarily stable equilibrium.

In Section 2, we introduce three-player games. In Section 3, we discuss adaptive dynamics 
with random matching of players. Results are presented in Section 4. Discussion follows in Section 5. 
In Appendix A, we present a tree representation of stationary states of irreducible Markov chains. 
Some more technical proofs are given in Appendix B.

\newtheorem{theo}{Theorem}
\newtheorem{defi}{Definition}
\newtheorem{hypo}{Hypothesis}
\newtheorem{example}{Example}
\newtheorem{corollary}{Corollary}
\newtheorem{lemma}{Lemma}
\newtheorem{prop}{Proposition}
 
\section{Three-player games}

To characterize a game-theoretic model, one has to specify the set of players, 
strategies they have at their disposal and payoffs they receive. The payoff 
of any player depends not only on his strategy but also on strategies 
played by his opponents. We will discuss symmetric three-player games
with two strategies. In symmetric games, all players
assume the same role in the game. It is enough therefore 
to specify payoffs of one player. It is given by two matrices:

\begin{equation}
U=\left( \left( 
\begin{array}{cc}
u_{111} & u_{121} \\ 
u_{211} & u_{221}
\end{array}
\right) ,\left( 
\begin{array}{cc}
u_{112} & u_{122} \\ 
u_{212} & u_{222}
\end{array}
\right) \right), 
\end{equation}

where $u_{ijk}$, $i,j,k =A,B$ is a payoff of the first (row) player when he plays 
the strategy $i$ when the second (column) player plays the strategy $j$, 
and the third (matrix) player plays $k$.

For example, if we assume that in the three-player stag-hunt game, a stag is worth, say, 6 units
and a hare 1 unit and one needs all three hunters to get a stag and share a reward, 
then the payoffs are given by  

\begin{equation}
U_{1} =
\left(
\left( \begin{array}{cc}
2 & 0 \\
1 & 1 \\
\end{array} \right)
\left( \begin{array}{cc}
0 & 0 \\
1 & 1 \\
\end{array} \right)
\right)
\end{equation}

In general we assume that the payoff of any player depends only on his strategy and numbers 
of players playing different types of strategies. We consider the following payoffs:

\begin{equation}
U_2 =
\left(
\left( \begin{array}{cc}
a_1 & b_1 \\
a_2 & b_2 \\
\end{array} \right)
\left( \begin{array}{cc}
b_1 & c_1 \\
b_2 & c_2 \\
\end{array} \right)
\right)
\end{equation}

Let $a=a_1 - a_2, b=b_2 - b_1$, and $c=c_2 - c_1$.
We will discuss all three classes of generic symmetric three-player games;
see (Bukowski and Mi\c{e}kisz, 2004) for the  complete classification.

In the first class, we have $a>0$ and $c>0$ and therefore two  
evolutionarily stable strategies: $A$ and $B$. 

A mixed strategy is a probability distribution on the set of pure strategies. 
It can be represented by $x, 0 \leq x \leq 1$, 
that is a probability of playing the first strategy $A$.  
The payoffs of $A$ and $B$ against $x$ are given 
by the expected values,  $a_{1}x^{2}+2b_{1}x(1-x)+c_{1}(1-x)^{2}$ 
and $a_{2}x^{2}+2b_{2}x(1-x)+c_{2}(1-x)^{2}$ respectively. Any $x^{*}$ 
for which the above two expected values are equal is a mixed Nash equilibrium.
$x^{*}$ can also be interpreted as a fraction of the population playing 
the first strategy in equilibrium. 

In the first class of games, in addition to two pure Nash equilibria, there exists also 
an unstable mixed Nash equilibrium, $x^{*}=\frac{(b-c)+\sqrt{b^2+ac}}{a+2b-c}$ if $b \neq (c-a)/2$
and $x^{*}=\frac{c}{a+c}$ otherwise.

In our second class of games, we have $a>0$ and $c<0$ (analogous results hold for $a<0<c$). 
Then $A$ is the only pure evolutionarily stable strategy. If $b > \sqrt{|ac|}$, then we have in addition 
a mixed evolutionarily stable strategy, $y^{*}=\frac{(b-c)-\sqrt{b^2+ac}}{a+2b-c}$,
and an unstable mixed Nash equilibrium $x^{*}$.

In the third class of games, $a<0$ and $c<0$ and there is 
a unique mixed evolutionarily stable strategy $x^{*}$.

Below we study stability of above evolutionarily stable strategies in adaptive stochastic dynamics
with random matching of players.

\section{Adaptive dynamics with random matching of players}

We consider a finite population of $n$ individuals
who have at their disposal one of two strategies:
$A$ and $B$. They are randomly matched in triples (we assume that $n$ is divisible by $3$)
to play a three-player symmetric game with payoffs given by $U_{2}$.

At every discrete moment of time, $t=1,2,...$, the state of the population
is described by the number of individuals, $z_{t}$, playing $A$. 
Formally, by the state space we mean the set $\Omega=\{z, 0 \leq z\leq n\}.$
Due to the random nature of matching, average payoffs 
of strategies depend on the realization of two random variables: 
$p_{t}$ - the number of triples in which there are exactly two $A$-players 
and $q_{t}$ - the number of triples with exactly one $A$-player. 
It follows that $\frac{z_{t} - 2p_{t} - q_{t}}{3}$ is the number of triples,
where all individuals play the first strategy. 

Now we will describe the dynamics of our system.
It consists of two components: selection and mutation.
The selection mechanism ensures that if the average payoff 
of a given strategy, $\pi_{i}(z_{t},p_{t},q_{t}), i=A,B$, 
at the time $t$, is bigger than the average payoff 
of the other one, then the number of individuals
playing the given strategy should increase in $t+1$. 
We assume that at any time period, each individual 
has a revision opportunity with a small positive probability $\tau$. 

Average payoffs are given by following expressions:

\begin{equation}
\pi_A(z_t, p_t, q_t) = \frac{a_{1}(z_t - 2p_t - q_t)+2b_{1}p_{t} +c_{1}q_{t}}{z_t},
\end{equation}
$$\pi_B(z_t, p_t, q_t) = \frac{a_{2}p_{t}+2b_{2}q_{t}+c_{2}(n-z_t-p_t-2q_t)}{(n-z_t)},$$
provided $0<z_{t}<n$. 

The selection dynamics is formalized in the following way:

\begin{equation}
z_{t+1} > z_{t} \hspace{2mm} if \hspace{2mm} \pi_{A}(z_{t},p_{t},q_{t}) > 
\pi_{B}(z_{t},p_{t},q_{t}),
\end{equation}
$$z_{t+1} < z_{t} \hspace{2mm} if \hspace{2mm} \pi_{A}(z_{t},p_{t},q_{t}) 
< \pi_{B}(z_{t},p_{t},q_{t}),$$
$$z_{t+1}= z_{t} \hspace{2mm} if \hspace{2mm} \pi_{A}(z_{t},p_{t},q_{t}) = 
\pi_{B}(z_{t},p_{t},q_{t}),$$
$$z_{t+1}= z_{t} \hspace{2mm} if \hspace{2mm} z_{t}=0  \hspace{2mm} 
or \hspace{2mm} z_{t}=n.$$

Now mutations are added. At every time period, each player who has 
a revision opportunity, instead of following the selection rule may adopt the other strategy
with a small probability $\epsilon$. It is easy to see, 
that for any two states of the population, there is a positive probability
of the transition between them in some finite number of time steps. 
We have therefore obtained an irreducible Markov chain with $n+1$ states. 
It has a unique stationary state (a probability mass function) 
which we denote by $\mu^{\epsilon}_{n}.$ For any $z \in \Omega$,
$\mu^{\epsilon}_{n}(z)$ is the frequency of visiting the state $z$ in the long run.
The following definition was first introduced by Foster and Young (1990).
\vspace{2mm}

\noindent {\bf Definition} $z \in \Omega$ is stochastically stable if 
$ \lim_{\epsilon \rightarrow 0}\mu_{n}^{\epsilon}(z)>0.$
\vspace{2mm}

\noindent In most cases below, there exists a state for which the above limit is actually equal to 1. 
In particular, unless we explicitly desribe the situation, if $z=n$ ($z=0$) 
is stochastically stable, then in the long run, in the limit of no mutations, 
all individuals play the strategy $A$ ($B$). We also say that respective strategies 
are stochastically stable.
 
\section{Stochastic stability of evolutionarily stable strategies}

We study here the stability of evolutionarily stable strategies of three-player games 
in described above adaptive stochastic dynamics with random matching of players.
We present our results and give some proofs. All remaining proofs are given in Appendix B. 
They are based on a certain tree representation of stationary states of irreducible Markov chains
(Freidlin and Wentzell, 1970 and 1984; see Appendix A). 
We assume that at any time period, each individual has a revision opportunity with a small
positive probability $\tau$. It follows that $z=0$ and $z=n$ are the only absorbing states. 
After a finite number of steps of the mutation-free dynamics we arrive 
at one of these two states and stay there forever - there are no other recurrence classes. 
Therefore to obtain a stationary state in the limit of no mutations, it is enough to count 
a number of mutations the population needs to evolve between these states. 
If one requires for example fewer mutations to evolve from $z=0$ to $z=n$ than from $z=n$ to $z=o$, 
then $z=n$ is stochastically stable. This means that in the long run, 
if the mutation level is sufficiently low, then almost all individuals play $A$. 
\vspace{2mm}

For games in the first class, without a loss of generality, we assume that $a_{1}> c_{2}$.
Under an additional condition, we have the following theorem (a proof is given in Appendix B).

\begin{theo}
If $a,c>0$, $a_{1}>b_{2},c_{2}$, and $n$ is sufficiently big, then $A$ is stochastically stable.
\end{theo}

In particular, we see that in the stag-hunt game with payoffs given by $U_{1}$,
the strategy $S$ is stochastically stable if the number of players is sufficiently large.
Observe that for the stag-hunt game with just three players, one needs three mutations 
to evolve from $z=0$ to $z=3$ and only one mutation to evolve from $z=3$ to $z=0$, 
hence $H$ is stochastically stable. We see that the stochastic stability depends 
on the number of players.

In general, we have the following theorem.

\begin{theo}
If $a,c>0$, $a_{1} > c_{2}$, $b_{1}<c_{2}$, and $(a_{1}(n-3)+c_{1})/(n-2) < b_{2} < a_{1}$, 
then $B$ is stochastically stable.
\end{theo}

\noindent {\bf Proof:} We have that $\pi_{B}(z=1) > \pi_{A}(z=1)$ and $\pi_{B}(z=n-1) < \pi_{A}(z=n-1)$
for sufficiently big $n$. For $z=n-2$, if two $B$-players are matched with one $A$-player, 
then $\pi_{B} > \pi_{A}$ and for $z=2$, for two possible matchings $\pi_{B} > \pi_{A}$.  
Hence the population needs only two mutations to evolve from $z=n$ to $z=0$ and three mutations 
to evolve from $z=0$ to $z=n.$ It follows from the tree representation of stationary states 
(see Appendix A) that $B$ is stochastically stable. $\Box$

It follows from Theorem 2 that if $b_{2}<a_{1}$ and $n<n^{*}=(3a_{1}-2b_{2}-c_{1})/(a_{1}-b_{2})$, 
then $B$ is stochastically stable. Observe that $n^{*}$ can be arbitrarily big. 
However, Theorem 1 tells us that $A$ is stochastically stable if $n$ is big enough. 
We see that when $n$ increases, the population undergoes a transition between its equilibria.   
\vspace{1mm}

Let us discuss now the case $a_{1}<b_{2}$. If $b_{1}<c_{2}$, then we may repeat the proof of Theorem 2
and obtain 

\begin{theo}
If $a,c>0$, $a_{1} > c_{2}, a_{1} < b_{2}$, and $b_{1}<c_{2}$, then $B$ is stochastically stable
if $n$ is sufficiently big.
\end{theo}

If $b_{1}> c_{2}$, then the population needs two mutations to evolve both from $z=n$ to $z=0$ 
and from $z=0$ to $z=n.$ We get the following theorem. 

\begin{theo}
If $a,c>0$, $a_{1} > c_{2}, a_{1}<b_{2}$, and $b_{1}>c_{2}$, then both $A$ and $B$ 
are stochastically stable if $n$ is sufficiently big.
\end{theo}

\vspace{2mm}
 
In games in the second class, the only pure evolutionarily stable strategy is stochastically stable.
Without a loss of generality, we assume that $a>0$ and $c<0$ and have the following theorem.

\begin{theo}
If $c<0<a$ and $n$ is sufficiently big, then $A$ is stochastically stable.
\end{theo}

\noindent {\bf Proof:} It is easy to see that for a big $n$, the population needs one mutation 
to evolve from $z=0$ to $z=n$ and at least two mutations to evolve from $z=n$ to $z=0$. $\Box$
\vspace{1mm}

However, we observe that for a fixed $n$, if $a_{1}<c_{2}$, $b_{1}$ is sufficiently small 
(negative and big in the absolute value) and $b_{2}$ is sufficiently  big, 
then $\pi_{B}(z) > \pi_{A}(z)$ for any $z \neq 0,n$ and all matchings. 
It follows that for such payoffs, $B$ is stochastically stable.

We will now show that for certain payoffs, although both $A$ and $B$ are stochastically stable, 
for a fixed and sufficiently small $\epsilon$, almost  $2/3$ of the population 
will play $B$ if $\tau$ is sufficiently small, that is when our dynamics is a small perturbation 
of the sequential one. The reason is that for small $\tau$, the probability that many players
adapt at the same time period is rather small - for $k$ players it is of order $\tau^{k}$
and moreover to evolve from $z=0$ to $z=n$, the population must pass through a large region
of $z's$, including $\{z, n/3 \leq z \leq 2n/3\}$, where  $\pi_{B}(z) > \pi_{A}(z)$ for any matching.

\begin{theo}
If $c<0<a$, $a_{1}<c_{2}$, $b_{1}<a_{2}$, $b_{2}>c_{1}$, and $n$ is sufficiently big, 
then for every $\delta>0$ and a sufficiently small $\epsilon$ there exists $\tau(\delta,\epsilon,n)$ such that 
if $\tau < \tau(\delta,\epsilon,n)$, then $\mu^{\epsilon,\tau}_{n}(z,z \leq n/3) >1-\delta$. 
\end{theo}

The proof is provided in Appendix B.
\vspace{2mm}

Finally we discuss the case $a<0$ and $c<0$. 

\begin{theo}
If $a,c<0$ and $n$ is sufficiently big, then both $A$ and $B$ are stochastically stable.
\end{theo}

\noindent {\bf Proof:} It is easy to see that for a big $n$, the population needs one mutation 
to evolve from $z=0$ to $z=n$ or from $z=n$ to $z=0$. 
\vspace{1mm}

As in the previous case, for a fixed $n$, if $a_{1}<c_{2}$, $b_{1}$ is sufficiently small 
and $b_{2}$ is sufficiently big, then $\pi_{B}(z) > \pi_{A}(z)$ for any $z \neq 0,n$
and any matching, and therefore $B$ is stochastically stable.

We also have the similar theorem concerning small $\tau$. 

\begin{theo}
If $a,c<0$, $a_{1}<c_{2}$, $b_{1}<c_{2}$, $b_{2}>a_{1}$, and $n$ is sufficiently big, 
then for every $\delta>0$ and sufficiently small $\tau$ there exists $\epsilon(\delta,\tau,n)$ such that 
if $\epsilon < \epsilon(\delta,\tau,n)$, then $\mu^{\epsilon,\tau}_{n}(0) >1-\delta$. 
\end{theo}

For a proof see Appendix B.

\section{Discussion}
Multi-player games naturally appear in many situations (see Broom {\em et al.,} 1997, 
for a list of biological examples involving leks and communal nests).
Here we discussed stochastic adaptive dynamics of populations of individuals 
playing three-player games. We assumed that players are matched with randomly chosen opponents. 
     
We considered all three classes of generic symmetric three-player games with two strategies. 
In the first class, there are two pure evolutionarily stable strategies.
We showed that only one is stochastically stable and therefore we resolved 
the problem of equilibrium selection. We also showed that the stochastic stability may depend
on the number of players. For some payoff parameters, when the number of players increases, 
the population may undergo a transition between its equilibria. 

In the second class of games, the unique pure evolutionarily stable strategy 
is also stochastically stable. However, for a fixed low level of mutation, 
when one considers very small probability of adaptation (when our dynamics is very close 
to a sequential one), then for certain payoff parameters, almost $2/3$ of all individuals 
will play in the long run the strategy which provides the higher payoff 
in the homogeneous population, although it is not evolutionarily stable. 

In the third class of games, both strategies are stochastically stable 
but they are not evolutionarily stable. This is a consequence of the fact that they are 
the only absorbing states and we take the limit of no mutations for a fixed number of players. 
If we first take the limit of an infinite number of players and then the limit of no mutations, 
then the long behavior may be different; see (Samuelson, 1997) for a discussion
on the order of taking different limits in evolutionary models.  

In our models we assumed that individuals are matched completely randomly.
We may allow some clustering in the population, either of geographic nature 
or following for example from the fact that $A$-players are more likely to meet other $A$-players. 
However, if any matching has a positive probability to occur, then our results will not change
qualitatively. This is a consequence of taking a limit of no mutations for a fixed number of players.
Stochastic stability of spatial three-player games with local interactions was studied in (Mi\c{e}kisz, 2004).
  
The stochastic stability concept involves the limit of no mutations for a fixed number of players.
However, for any arbitrarily low level of mutations, when the number of players is sufficiently big, 
then the long-run behavior of a population might be different. In fact, it was shown (Mi\c{e}kisz, 2005) 
that in the case of two-player games with two strategies, in the limit of the infinite number of players 
in the random matching model, the long run behavior of the population is similar to the one 
in the Kandori-Mailath-Rob model (Kandori {\em et al.}, 1993). The assumption  of the infinite number 
of players has the same effect as the limit of infinitely many matchings per time period. The second limit 
can be justified if the rate of playing games is much bigger than the rate of adaptation.
We expect the same situation in the case of three-player games. The stochastic stability 
in multi-player games in the limit of infinitely many matchings per time period 
was investigated by Kim (1996). 

In order to study the long-run behavior of stochastic population dynamics,  
we should estimate the relevant parameters to be sure what limiting procedures 
are appropriate in specific biological examples. 
\vspace{4mm}

\noindent {\bf Acknowledgments} JM would like to thank 
the Polish Committee for Scientific Research for a financial support
under the grant KBN 5 P03A 025 20.
\eject

\noindent {\bf Appendix A}
\vspace{2mm}

\noindent The following tree representation of stationary distributions 
of Markov chains was proposed by Freidlin and Wentzell (1970, 1984). 
Let $(\Omega,P)$ be an irreducible Markov chain with a state space 
$\Omega$ and transition probabilities given by $P^{\epsilon}: \Omega \times \Omega \rightarrow [0,1]$. 
It has a unique stationary distribution, $\mu^{\epsilon}$, also called a stationary state. 
For $X \in \Omega$, let an X-tree be a directed graph on $\Omega$ such that from every $Y \neq X$ 
there is a unique path to $X$ and there are no outcoming edges out of $X$. 
Denote by $T(X)$ the set of all X-trees and let 
\begin{equation}
q^{\epsilon}(X)=\sum_{d \in T(X)} \prod_{(Y,Y') \in d}P^{\epsilon}(Y,Y'),
\end{equation}
where the product is with respect to all edges of $d$. 
We have that
\begin{equation}
\mu^{\epsilon}(X)=\frac{q^{\epsilon}(X)}{\sum_{Y \in \Omega}q^{\epsilon}(Y)}
\end{equation}
for all $X \in \Omega.$

Let us assume now that after a finite number of steps of the mutation-free dynamics, i.e. $\epsilon=0$, 
we arrive at one of two absorbing states, say $X$ and $Y$, and stay there forever - there are 
no other recurrence classes. It follows from the tree representation that any state different from absorbing states 
has zero probability in the stationary distribution in the zero-mutation limit. 
Consider a dynamics in which $P^{\epsilon}(Z,W)$ for all $Z, W \in \Omega$, is of order $\epsilon^{m}$, 
where $m$ is the number of mutations involved to pass from $Z$ to $W$ or is zero. Then one has to compute
the minimal number of mutations, $m_{XY}$, needed to make a transition from the state $X$ 
to $Y$ and the number of mutations, $m_{YX}$, to evolve from $Y$ to $X$. 
$q(X)$ is of order $\epsilon^{m(YX)}$ and $q(Y)$ is of order $\epsilon^{m(XY)}$.  
It follows that if $m_{XY} < m_{YX}$, then $Y$ is stochastically stable and moreover 
$ \lim_{\epsilon \rightarrow 0}\mu^{\epsilon}(Y)=1.$    
\eject

\noindent {\bf Appendix B}
\vspace{2mm}

\noindent {\bf Proof of Theorem 1:}
\vspace{2mm}

\noindent We begin by examining in detail the mutation-free dynamics. 
First we show that if $a,c>0$, then $z=0$ and $z=n$ 
are the only absorbing states (for a sufficiently big $n$) 
of the mutation-free dynamics even in the sequential dynamics, 
i.e. when at any time period only one randomly chosen player 
can adapt to the environment. Let $U_{A}$ and $U_{B}$ be the basins of attraction 
of $z=n$ and $z=0$ respectively. Let us emphasize that 
because of the stochastic nature of matchings, they overlap substantially. 
Observe first that there exists $k^{*}$ such that if $k \geq k^{*}$, 
then $k \in U_{A}.$  If $k$ is divisible by $3$ and all $A$-players 
and $B$-players are matched within themselves, then $\pi_{A} > \pi_{B}.$ 
If $k=3m+1$ for some natural number $m$, then let two pairs of $A$-players 
be matched with $B$-players; if $k=3m+2$, then let two $A$-players 
be matched with one $B$-player. In both cases, all other $A$-players 
are matched within themselves. Now for a sufficiently big $k^{*}$ we have 

\begin{equation}
a_{1}(k^{*}-4)/k^{*} > a_{2}
\end{equation}
 
and therefore the inequality $\pi_{A} > \pi_{B}.$ 
Let us observe that for any $z<k^{*}$ and $n$ big enough,
if all $A$-players are matched with two $B$-players, then $\pi_{A} < \pi_{B}.$ 
We have shown that $U_{A} \cup U_{B}= \Omega.$ 
Moreover, there are no other recurrence classes.

Assume now, without a loss of generality, that all payoffs are positive.
Let us notice that for $z=n-k^{*}$, if

\begin{equation}
a_{1}(n-2k^{*})/(n-k^{*}) > max\{a_{2}, b_{2}, c_{2}\} \equiv f
\end{equation} 
which is true if 

\begin{equation}
n>k^{*}(2a_{1}-f)/(a_{1}-f),
\end{equation}
then  $\pi_{A}> \pi_{B}$ for any matching. It follows that the size of $U_{B}$ is at most $n-k^{*}-1$ 
and that of $U_{A}$ at least $n-k^{*}.$ Hence the population needs 
at least $k^{*}+1$ mutations to evolve from $z=n$ to $z=0$ 
and at most $k^{*}$ mutations to evolve from $z=0$ to $z=n.$ 
It follows from the tree representation of stationary states (see Appendix A)
that $z=n$ is stochastically stable. 
\vspace{2mm}

\noindent {\bf Proof of Theorem 6:} 
\vspace{2mm}

For every $z=k$ with $k$ divisible by $3$, 
if all $B$-players are matched within themselves, 
then  $\pi_{B} > \pi_{A}$ and therefore the number of $B$-players 
increases in the mutation-free dynamics. It is also true if $k$ is not too small and not too big
(independent of $n$) and only one or two $B$-players are matched with $A$-players. 
We have that $\pi_{B}(z=1) < \pi_{A}(z=1)$ and $\pi_{B}(z=n-1) < \pi_{A}(z=n-1).$ 
However, for other small and big $k's$ (depending on $n$), 
there is a positive probability that $\pi_{B}(z=k)>\pi_{A}(z=k).$ 
Note that from the assumptions of the theorem follows
that $b_{1} < a_{2}, b_{2}, c_{2}$ and $b_{2}>a_{1}, b_{1}, c_{1}$.
Therefore it is enough that all $A$-players are matched with one $B$-player 
and one $A$-player for small $k's$ and all $B$-players are matched 
with one $A$-player and one $B$-player for big $k's$. 
We get that there is a positive probability that 
$\pi_{B}(z)> \pi_{A}(z)$ if $z \neq 0,1,n-1,n$. 

If $k=n/2$ and there are the same number of triples with two $A$-players 
and one $B$-player and two $B$-players and one $A$-player, then   
$\pi_{B}> \pi_{A}$ because $2b_{1}+c_{1} < a_{2} + 2b_{2}$. 
One may check that the same conclusion follows if $2n/3$ or more individuals 
are matched in this way and the rest are either $A$ or $B$-players matched within themselves.
If $k =n/3$ and all $A$-players are matched with two $B$-players or if 
$k=2n/3$ and all $B$-players are matched with two $A$-players then again
$\pi_{B} > \pi_{A}$. It follows that if $z=k$ and $n/3 \leq k \leq 2n/3$, then 
$\pi_{B} > \pi_{A}$ for all possible matchings.

Now for a sufficiently small $\tau$, one considers only trees of order $\tau^{n}$, 
where at any time period, only one randomly chosen individual may adapt 
to the invironment - our dynamics is then a small perturbation of the sequential one. 
Among these trees there are $(z=k)$-trees with $0 \leq k \leq n/3$ which are of order $\epsilon^{2}$. 
Observe that $\{z\in \Omega, z=k, 0<k<n/3\}$ is a recurrence class for the sequential dynamics.
The thesis follows for sufficiently small $\epsilon$ and $\tau$.  
\vspace{2mm}

\noindent {\bf Proof of Theorem 8:} 
\vspace{2mm}

Again, as in the previous case, if $k$ is not too small and not too large (independent of $n$), 
then there is a positive probability that $\pi_{B}(z=k) > \pi_{A}(z=k)$ for any matching. 
It is easy to see that if $b_{2} > a_{1}$, then $\pi_{B}(z) > \pi_{A}(z)$ for all matchings
if $z=k$ and $k$ is sufficiently big; $k>\alpha n$ for some $\alpha>0$.  
If $c_{2} > b_{1}$, then there is a positive probability that $\pi_{B}(z)>\pi_{A}(z)$ 
for small $z\neq 0,1$ $z=k$, $k<\beta n$ for some $\beta>0$. 
It follows that there is a positive probability that $\pi_{B}(z) > \pi_{A}(z)$ for any $z \neq 0,1,n$. 

Now for a sufficiently small $\epsilon$, one considers only $(z=0)$-trees with just one mutation.
Among these trees there is one which is of order $\tau^{n+2}$. In this tree, 
the number of individuals playing $A$ decreases by one, from $z=n$ until $z=3$, 
then it jumps to $z=1$, then it increases to $z=2$, and finally it jumps down to $z=0$. 
Such tree, for a sufficiently small $\tau$, is the main contribution to $q(z=0)$.  
All $(z=n)$-trees with just one mutation are of order at least $\tau^{n(1+\alpha)}$. 
The thesis follows for sufficiently small $\tau$ and $\epsilon$.  
\vspace{2mm}

\noindent {\bf Bibliography}
\vspace{2mm}
 
\noindent Broom, M., C. Cannings, and G. T. Vickers (1997).
Multi-player matrix games. {\em Bull. Math. Biology} {\bf 59}: 931-952.
\vspace{2mm}

\noindent Bukowski, M. and J. Mi\c{e}kisz (2004). 
Evolutionary and asymptotic stability in multi-player games with two strategies.
{\em Int. J. Game Theory} {\bf 33}: 41-54. 
\vspace{2mm}

\noindent Foster, D. and P. H. Young (1990).
Stochastic evolutionary game dynamics.
{\em Theoretical Population Biology} {\bf 38}: 219-232.
\vspace{2mm}

\noindent Freidlin, M. and A. Wentzell (1970). 
On small random perturbations of dynamical systems. {\em Russian Math. Surveys}
{\bf 25}: 1-55.
\vspace{2mm}

\noindent Freidlin, M. and A. Wentzell (1984).
{\em Random Perturbations of Dynamical Systems.} New York: Springer Verlag.
\vspace{2mm}

\noindent Hofbauer, J., P. Schuster, and K. Sigmund (1979).
A note on evolutionarily stable strategies and game dynamics.
{\em J. Theor. Biol.} {\bf 81}: 609-612.
\vspace{2mm}

\noindent Hofbauer, J. and K. Sigmund (1998). {\em Evolutionary Games and Population Dynamics.} 
Cambridge: Cambridge University Press.
\vspace{2mm}
 
\noindent Hofbauer, J. and K. Sigmund (2003). Evolutionary game dynamics. 
{\em Bulletin AMS} {\bf 40}: 479-519. 

\noindent Kandori, M., G. J. Mailath, and R. Rob (1993).
Learning, mutation, and long-run equilibria in games. 
{\em Econometrica} {\bf 61}: 29-56. 
\vspace{2mm}

\noindent Kim, Y. (1996). Equilibrium selection in n-person 
coordination games. {\em Games Econ. Behav.} {\bf 15}: 203-277.
\vspace{2mm}

\noindent Maynard Smith, J. and G. R. Price (1973). The logic of animal conflicts. 
{\em Nature} {\bf 246}: 15-18.
\vspace{2mm}

\noindent Maynard Smith, J. (1982). {\em Evolution and the Theory of Games}.
Cambridge: Cambridge University Press.
\vspace{2mm}

\noindent Mi\c{e}kisz, J. (2005). 
Equilibrium selection in evolutionary games with random matching of players.
{\em J. Theor. Biol.} {\bf 232}: 47-53.
\vspace{2mm}

\noindent Mi\c{e}kisz, J. (2004). 
Stochastic stability in spatial three-player games.
{\em Physica A} {\bf 343}: 175-184.
\vspace{2mm} 

\noindent Robson, A. and F. Vega-Redondo (1996).
Efficient equilibrium selection in evolutionary games with random matching. 
{\em J. Econ. Theory} {\bf 70}: 65-92.
\vspace{2mm}

\noindent Samuelson, L. (1997). {\em Evolutionary Games and Equilibrium Selection}.
Cambridge: MIT Press.
\vspace{2mm}

\noindent Taylor, P. D. and L. B. Jonker (1978). Evolutionarily stable
strategy and game dynamics. {\em Math. Biosci.} {\bf 40}: 145-156.
\vspace{2mm}

\noindent Vega-Redondo, F. (1996). {\em Evolution, Games, and Economic Behaviour.}
Oxford: Oxford University Press.
\vspace{2mm}

\noindent Weibull, J. (1995). {\em Evolutionary Game Theory.} Cambridge: MIT Press.
\vspace{2mm}

\noindent Zeeman, E. (1981). Dynamics of the evolution of animal conflicts.
{\em J. Theor. Biol.} {\bf 89}: 249-270.

\end{document}